\documentclass[%
 reprint,
superscriptaddress,
 amsmath,amssymb,
 aps,
floatfix,
]{revtex4-1}

\usepackage{graphicx}
\usepackage{dcolumn}
\usepackage{bm}

\usepackage{color}

\usepackage{ulem} 

\begin{document}


\title{Syntheses and First-Principles Calculations of the Pseudobrookite Compound AlTi$_2$O$_5$}

\author{Takami Tohyama}
\email[e-mail:]{tohyama@rs.tus.ac.jp}
\affiliation{Department of Applied Physics, Tokyo University of Science, Tokyo 125-8585, Japan}

\author{Riku Ogura}
\affiliation{Department of Applied Physics, Tokyo University of Science, Tokyo 125-8585, Japan}

\author{Kensuke Yoshinaga}
\affiliation{Department of Applied Physics, Tokyo University of Science, Tokyo 125-8585, Japan}

\author{Shin Naito}
\affiliation{Department of Applied Physics, Tokyo University of Science, Tokyo 125-8585, Japan}

\author{Nobuaki Miyakawa}
\affiliation{Department of Applied Physics, Tokyo University of Science, Tokyo 125-8585, Japan}

\author{Eiji Kaneshita}
\affiliation{National Institute of Technology, Sendai College, Sendai 989-3128, Japan}

\date{\today}
             

\begin{abstract}

We synthesize a new titanium-oxide compound AlTi$_2$O$_5$ with pseudobrookite-type structure. The formal valence of Ti is $3.5+$ that is the same as the Magn\'eli compound Ti$_4$O$_7$ and the spinel compound LiTi$_2$O$_4$. AlTi$_2$O$_5$ exhibits insulating behavior in resistivity. Using experimentally determined crystal structure, we perform the first-principles calculations for the electronic structures. Experimentally suggested random distribution of Al and Ti is not the origin of the insulating behavior. The Fermi surfaces for nonrandom AlTi$_2$O$_5$  show cylindrical shapes reflecting a layered structure, which indicates a possible nesting-driven order. Constructing a tight-binding model from the first-principles calculations, we calculate the spin and charge susceptibilities using the random phase approximation. We suggest possible charge-density-wave state forming Ti$^{3+}$ chains separated from each other by Ti$^{4+}$ chains, similar to the low-temperature phase of Ti$_4$O$_7$.

\end{abstract}
\maketitle

\section{Introduction}
Titanium sub-oxides Ti$_n$O$_{2n-1}$ ($n\ge 4$) are known as the Magn\'eli phase. Among them, Ti$_4$O$_7$ is a mixed valence compound with Ti$^{3.5+}$.  In the high-temperature phase above 154~K, there is a clear metallic Fermi edge~\cite{Taguchi2010}. In the low-temperature phase below 142~K, it is well established that Ti$_4$O$_7$ shows charge ordered chains of Ti$^{3+}$ ($3d^1$) separated from each other by Ti$^{4+}$ ($3d^0$) chains~\cite{Marezio1973} with a clear insulating gap of $\sim$100~meV~\cite{Taguchi2010}. The $3d$ electrons are believed to be localized by forming bipolaron with Ti$^{3+}$-Ti$^{3+}$ pairs~\cite{Lakkis1976}. Such a charge ordering of Ti$^{3+}$ and Ti$^{4+}$ is also expected in other titanium oxides with Ti$^{3.5+}$. Although the spinel compound LiTi$_2$O$_4$ has the formal valence of Ti$^{3.5+}$, it becomes superconducting~\cite{Johnston1973}.  

\begin{figure}[tb]
\begin{center}
\includegraphics[width=0.48\textwidth]{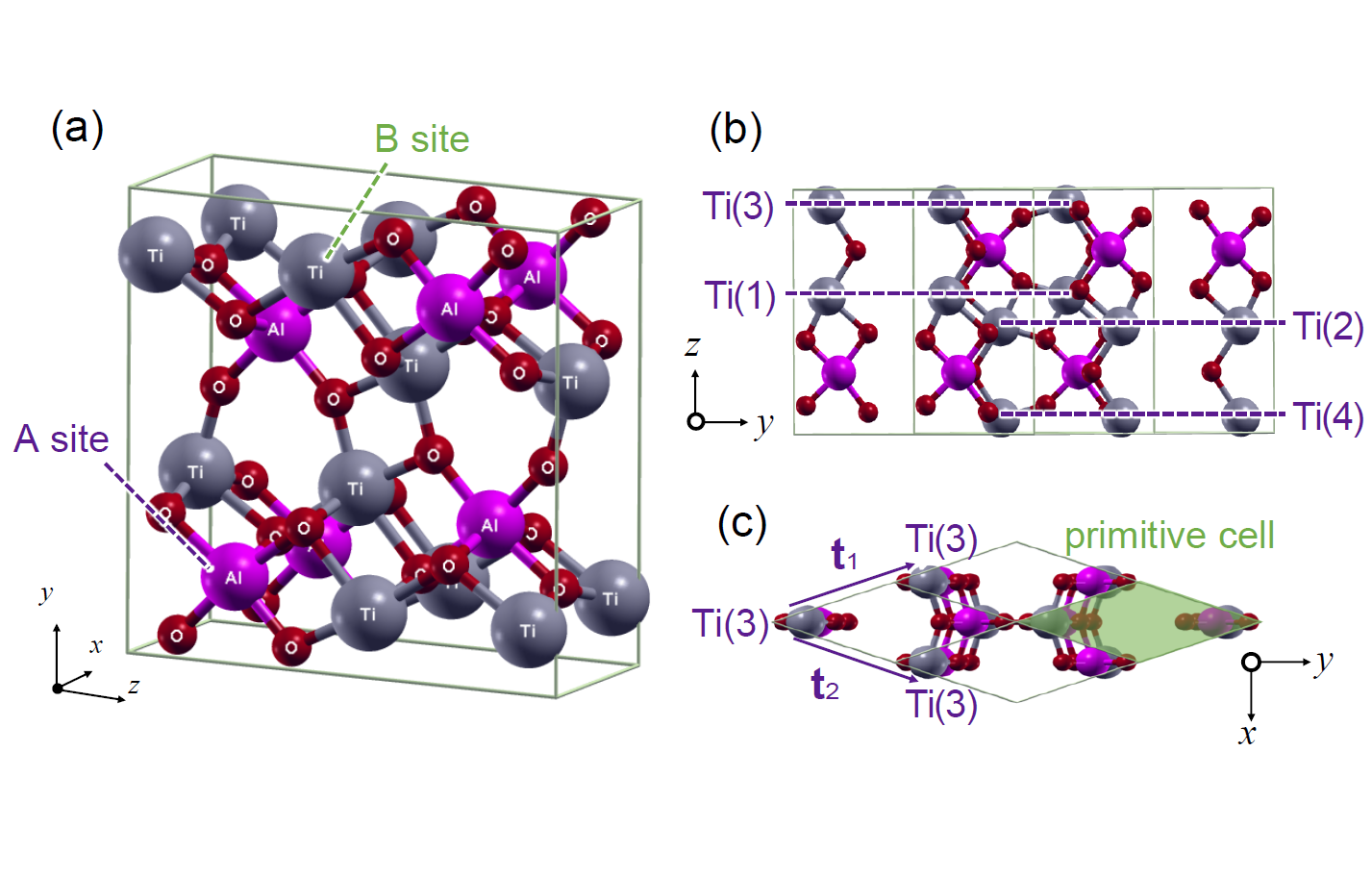}
\caption{(a) Crystal structure of pseudobrookite compound AlTi$_2$O$_5$ without randomness between Al and Ti. Al and Ti are assumed to occupy the A and B sites, respectively. There are four Ti site in the unit cell. (b) The projection onto the $b$--$c$ plane. Ti(1), Ti(2), Ti(3), and Ti(4) form each other a layered structure as shown by the dotted lines. (c) The projection onto the $a$-$b$ plane. $\mathbf{t}_1$ and $\mathbf{t}_2$ represent the translational vector. The green region corresponds to the primitive unit cell.}
\label{figST}
\end{center}
\end{figure}

In this paper, we synthesize a new titanium-oxide compound AlTi$_2$O$_5$. The formal valence of Ti is $3.5+$ that is the same as Ti$_4$O$_7$. AlTi$_2$O$_5$ has pseudobrookite-type structure as shown in Fig.~\ref{figST}(a)], where Al (Ti) is assumed to occupy the A (B) site. In the synthesized sample, Al and Ti are suggested to be distributed on both the A and B sites, although more than 60~\% of the B site is occupied by Ti. AlTi$_2$O$_5$ is insulating, showing resistivity that increases slightly with decreasing temperature from the room temperature and shows huge enhancement below 120~K. Although there is no signature of phase transition, the enhancement suggests a drastic change of electronic states at low temperature. Since Ti$^{3.5+}$ suggests a metallic behavior, it is important to elucidate theoretically how the insulating behavior emerges in AlTi$_2$O$_5$. We then investigate the electronic structure of AlTi$_2$O$_5$ and predict possible mechanism of its insulating behavior. The electronic structure is examined by the first-principles calculations. Because of an approximate layered structure in AlTi$_2$O$_5$, we find cylindrical Fermi surfaces for nonrandom distribution of Al and Ti, indicating a possible nesting-driven order. However, there is neither insulating nor magnetic solution in the first-principles calculations. Even for the randomly distributed cases, a super-cell calculation shows a metallic solution. In order to take a crucial insight for the origin of insulating behavior, we perform the calculation of spin and charge susceptibilities using random phase approximation (RPA) for a tight-binding model of the nonrandom AlTi$_2$O$_5$ constructing from the first-principles calculation. We find that the charge susceptibility shows strong enhancement at the wave vectors close to nesting conditions. Their positions depend on the choice of Coulomb interactions. One of the possible wave vectors corresponds to a charge-density-wave (CDW) state similar to the low-temperature phase of Ti$_4$O$_7$. The enhancement is expected to be due to nesting properties of the Fermi surfaces in addition to the presence of the van Hove singularity near the Fermi level. These results will give a hint for the mechanism of insulating behavior in AlTi$_2$O$_5$.     

This paper is organized as follows. Sample preparation and characterization of AlTi$_2$O$_5$ are given in Sec.~\ref{Sec2}. The first-principles calculations of the electronic structures are performed in Sec.~\ref{Sec3}, where both nonrandom and random structures are taken into account. In Sec.~\ref{Sec4}, we calculate spin and charge susceptibility within RPA based on a tight-binding Hamiltonian derived from the nonrandom band structure of AlTi$_2$O$_5$. Orbital-dependent charge fluctuations at possible CDW states are also shown. Finally, a summary is given in Sec.~\ref{Sec5}.

\section{Sample preparation and characterization}
\label{Sec2}
Polycrystalline AlTi$_2$O$_5$ crystals were prepared by the solid-state reaction method as follows:
$\mathrm{Al}_2\mathrm{O}_3+2\mathrm{TiO}_2+\mathrm{Ti}_2\mathrm{O}_3\rightarrow 2\mathrm{AlTi}_2\mathrm{O}_5$.
The starting materials were powders of Al$_2$O$_3$ (Furuuchi Chemical Co. 99.9~\%), rutile-type TiO$_2$ (Kojundo Chemical Laboratory Co., Ltd 99.9~\%) and Ti$_2$O$_3$ (Furuuchi Chemical Co.99.9~\%). The mixture of the starting compounds was set at the molar ratio of $\mathrm{Al}_2\mathrm{O}_3 : \mathrm{TiO}_2 : \mathrm{Ti}_2\mathrm{O}_3 = 1 : 2 : 1$, mechanically ground in an agate mortar for 2 h, and pressed into pellets. The pellets were sealed in a quartz ampoule under a vacuum of $\sim 10^{-2}$ Torr and sintered at $1300\,^{\circ}\mathrm{C}$ for 3 days, and the ampule was quenched in water. 

\begin{figure}[tb]
\begin{center}
\includegraphics[width=0.4\textwidth]{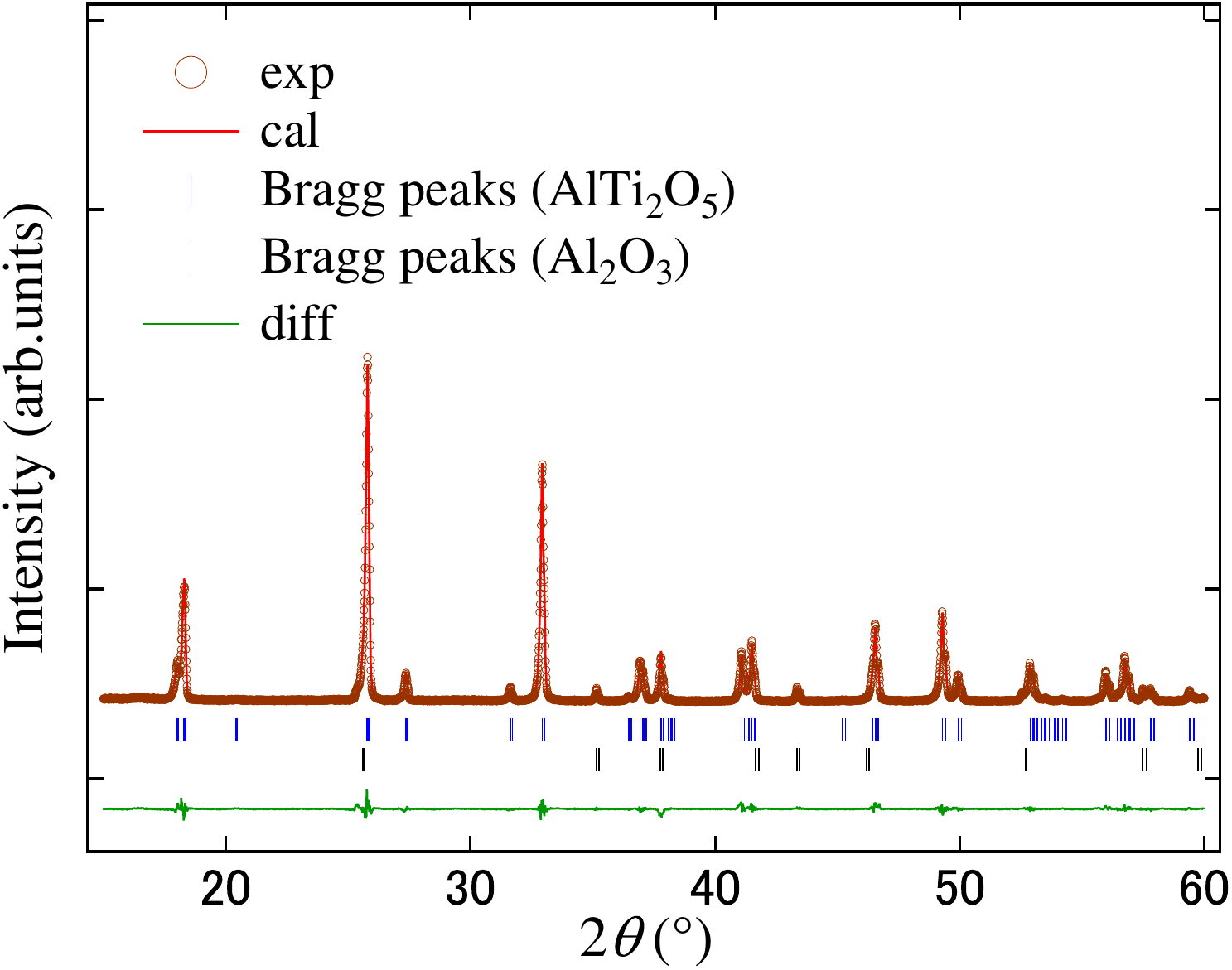}
\caption{Rietveld refinement of PXRD pattern of pseudobrookite-AlTi$_2$O$_5$. The brown open circles represent the experimental data, the red solid line is the calculated pattern, the green solid line is the difference between the experimental and calculated values, and the blue and black vertical bars stand at calculated $2\theta$ angles for the Bragg peaks of AlTi$_2$O$_5$ (top) and Al$_2$O$_3$ (bottom), respectively.}
\label{figRietvelt}
\end{center}
\end{figure}

The products of AlTi$_2$O$_5$ were characterized by the powder X-ray diffraction (PXRD) using a RINT2500V diffractometer (Rigaku, Japan) in flat plate geometry with CuK$_\alpha$ radiation ($\lambda = 1.5418$~\AA) at 40~kV and 200~mA. PXRD data were typically collected in the range $15\,^{\circ} \leq 2\theta \leq 60\,^{\circ}$ with a step rate of $0.01\,^{\circ}$/sec. Rietveld refinement was performed using the RIETAN-FP software package~\cite{Izumi2007}. Figure~\ref{figRietvelt} shows the PXRD result for the polycrystalline sample. The diffraction pattern indicates pseudobrookite-AlTi$_2$O$_5$ as a dominant phase. However, there are two additional weak diffraction peaks from Al$_2$O$_3$ at $2\theta\sim 35.2\,^{\circ}$ and $\sim 43.4\,^{\circ}$. They were excluded from the Rietveld refinement for pseudobrookite-AlTi$_2$O$_5$, where isostructural Al$_2$TiO$_5$ was used as a reference model. The resulting refined structure is consistent with the orthorhombic crystal symmetry (space group: $C_\mathrm{mcm}$, No. 63). In Fig.~\ref{figRietvelt}, the green solid line is the difference between the experimental and calculated values, and the blue and black vertical bars are calculated $2\theta$ angles for the Bragg peaks of AlTi$_2$O$_5$ (top panel in Fig.~\ref{figRietvelt}) and Al$_2$O$_3$ (bottom panel in Fig.~\ref{figRietvelt}), respectively. We obtain reasonable reliability factor values of $R_\mathrm{wp}=9.436$~\%, $R_\mathrm{e}=6.634$~\% and $S=1.422$, and the refined lattice parameters of AlTi$_2$O$_5$ as $a=3.69363(5)$~\AA, $b=9.6894(1)$~\AA, and $c=9.84756(1)$~\AA. The refinement results, i.e., the crystallographic data of pseudobrookite-AlTi$_2$O$_5$, are summarized in Table~\ref{table1}, where the $x$, $y$, and $z$ exhibit the atomic coordinates and $g_i(S)$ represents the occupation rates with $i=$~Al, Ti, O and $S=$~A, B sites. The $g_\mathrm{O}$ value of oxygen is fixed as 1, while the occupation rate of Al and Ti in the A- and B-sites suggests possible random distribution of Al and Ti in the A site and B site. If there is no randomness, the B site is expected to be occupied by only Ti, while in the refined data 36~\% of the B site is occupied by Al. 

\begin{figure}[tb]
\begin{center}
\includegraphics[width=0.4\textwidth]{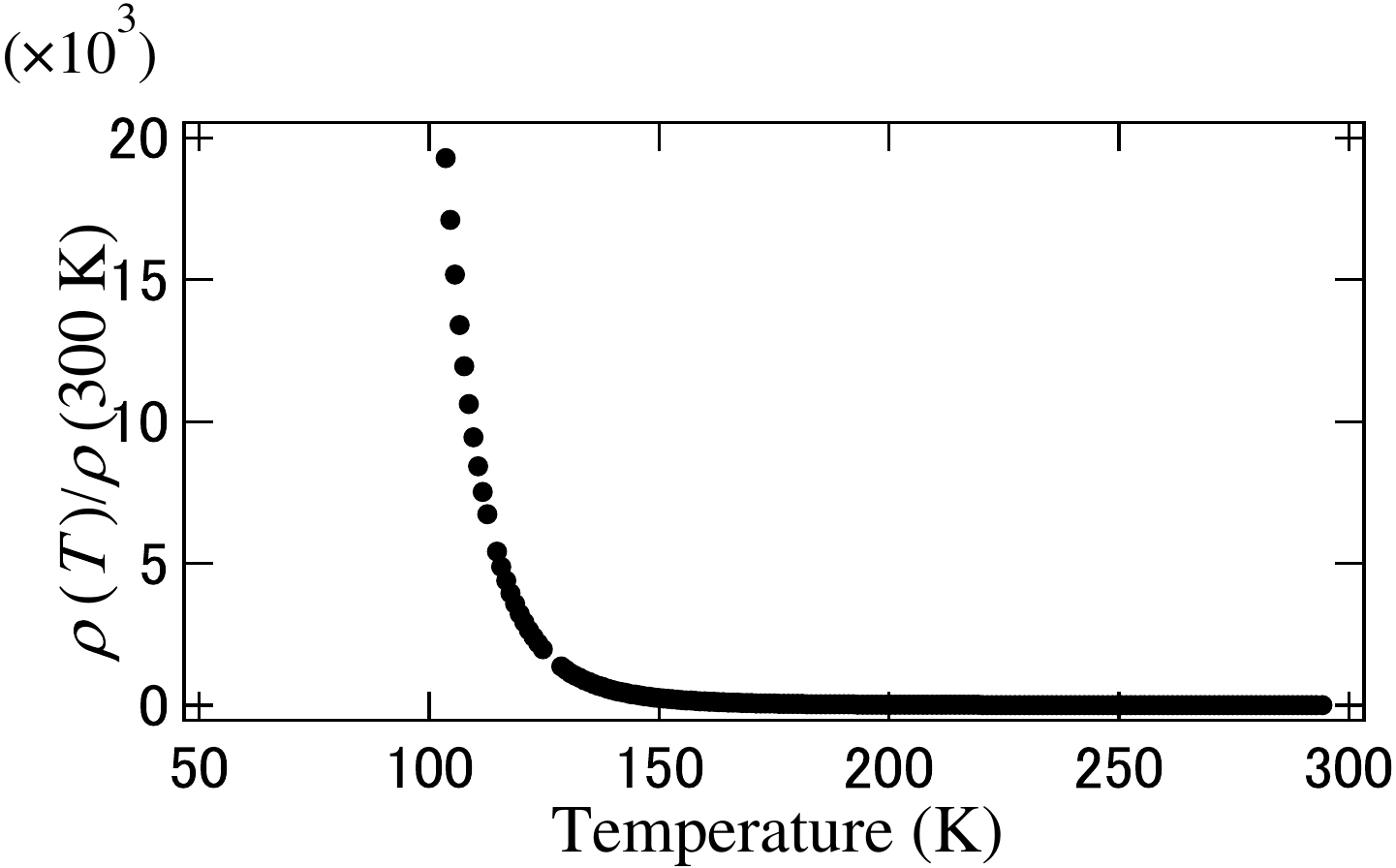}
\caption{Temperature dependence of the resistivity for AlTi$_2$O$_5$.}
\label{figResistivity}
\end{center}
\end{figure}

The electrical resistivity was measured with the conventional four-probe method. Temperature dependence of electrical resistivity$\rho(T)$ measured on polycrystalline AlTi$_2$O$_5$ is shown in Fig.~\ref{figResistivity}. The $\rho(T)$ shows the insulating behavior with $d\rho/dT<0$ in the range of $T<300$~K and a rapid increase below $T=120$~K, indicating a change of electronic states at low temperature. This insulating behavior did not change by the post-annealing in vacuum at $T<1000\,^{\circ}\mathrm{C}$. 

\begin{table}[tb]
\caption{\label{table1}
Crystallographic data of AlTi$_2$O$_5$ (space group: $C_\mathrm{mcm}$, No.63). Wyckoff positions (W.p.), atomic coordinates, occupancy $g_i$ for atom $i=$Al, Ti, and O, isotropic displacement parameters $B$ in unit of \AA$^2$.  Lattice constants are $a=3.69363(5)$~\AA, $b=9.6894(1)$~\AA, and $c=9.84756(1)$~\AA. The volume of the unit cell is $V_\mathrm{cell}= 352.434(8)$~\AA$^3$.}
\begin{ruledtabular}
\begin{tabular}{ccccccc}
\textrm{$i$(site)}&
\textrm{W.p.}&
\textrm{$x$}&
\textrm{$y$}&
\textrm{$z$}&
\textrm{$g_i$}&
\textrm{$B$}\\
\colrule
Al(A) & 4c & 0 & 0.1912(2) & 1/4 & 0.277(5) & 0.5\\
Ti(A) & 4c & 0 & 0.1912 & 1/4 & 0.7229 & 0.5\\
Al(B) & 8f & 0 & 0.1331(2) & 0.5598(1) & 0.3614 & 0.5\\
Ti(B) & 8f & 0 & 0.1331 & 0.5598 & 0.6386 & 0.5\\
O(1) & 4c & 0 & 0.7548(6) & 1/4 & 1 & 0.9\\
O(2) & 8f & 0 & 0.0470(3) & 0.1194(4) & 1 & 0.9\\
O(3) & 4c & 0 & 0.3160(4) & 0.0701(3) & 1 & 0.9\\
\end{tabular}
\end{ruledtabular}
\end{table}

\section{First-principles calculations}
\label{Sec3}
\subsection{Nonrandom case}
\label{Sec2-1}
AlTi$_2$O$_5$ shows a pseudobrookite-type structure with the space group $C_\mathrm{mcm}$.  For simplicity, we here assume that Al and Ti occupy the A and B sites, respectively, as shown in Fig.~\ref{figST}(a), where the translational vectors $\mathbf{t}_1=(a/2,b/2,0)$, $\mathbf{t}_2=(a/2,-b/2,0)$, and $\mathbf{t}_3=(0,0,c)$. There are four Ti sites denoted by Ti(1), Ti(2), Ti(3), and Ti(4) in the primitive unit cell. Each Ti site forms a layered structure along the $c$ direction, resulting in four layers [see Fig.~\ref{figST}(b)]. This indicates a pseudo two-dimensional (2D) electronic structure in the $a$--$b$ plane.  

The electronic structure was calculated using density functional theory based on the Perdew, Burke, Ernzerhof (PBE) generalized gradient approximation (GGA)~\cite{Perdew1996}. We used the general potential linearized augmented planewave (LAPW) method including local orbitals~\cite{Sjostedt2000} as implemented in the WIEN2k package~\cite{Blaha2001}. The muffin-tin radii were 1.58, 1.87, and 1.69 Bohr for Al, Ti, and O, respectively. We employed atomic coordinates listed in Table~\ref{table1}.  A 12$\times$12$\times$12 $k$-point grid was used for the Brillouin zone (BZ) integration.

\begin{figure}[tb]
\begin{center}
\includegraphics[width=0.48\textwidth]{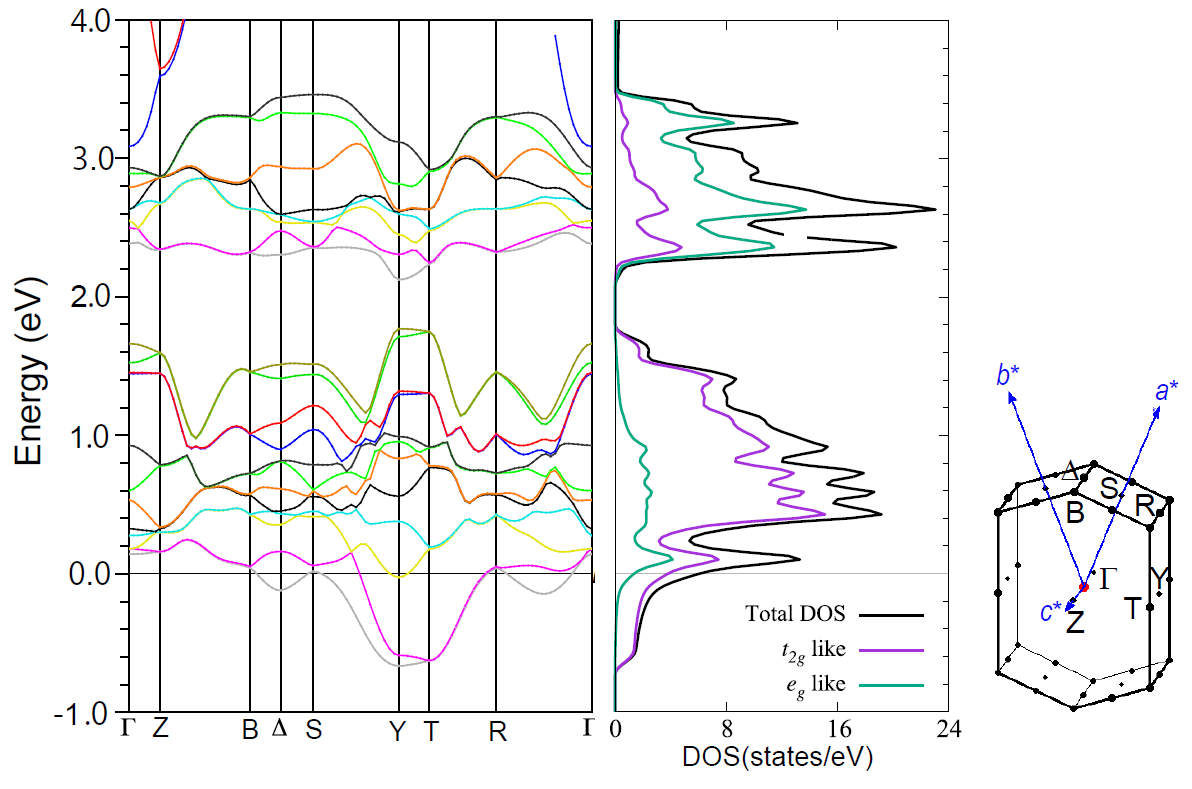}
\caption{Calculated band structure (left) along the symmetric lines in the BZ (right) and DOS (middle) of the pseudobrookite-type AlTi$_2$O$_5$ with experimental lattice parameters. The DOS from the $t_{2g}$-like and $e_g$-like orbitals in Ti atom are plotted together with the total DOS.}
\label{figBS}
\end{center}
\end{figure}

The calculated band structure and the corresponding electronic density of states (DOS) are shown in Fig.~\ref{figBS}.  The band in the energy range from $-$0.8~eV to 3.6~eV is predominantly composed of Ti $3d$ orbitals, and it is separated to the two regions, $-$0.8~eV to 1.8~eV and 2.3~eV to 3.6~eV, due to the crystal field induced by octahedral coordinates of eight oxygen ions around a Ti ion. The former region has 12 bands with $t_{2g}$-like components dominated by $d_{yz}$, $d_{xz}$, and $d_{x^2-y^2}$, while the latter region has 8 bands with $e_g$-like components dominated by $d_{xy}$ and $d_{3z^2-r^2}$.  The Fermi level $E_\mathrm{F}$ whose energy is 0~eV is located on the $t_{2g}$-like band, since the formal valence of Ti is 3.5+ ($3d^{0.5}$). The DOS at $E_\mathrm{F}$ is away from either a peak or a valley of the DOS. Therefore, magnetic state is unstable in our DFT calculation. In fact, a GGA+U calculation with on-site Coulomb energy $U=5$~eV did not show any magnetic solution. We note that the signature of quasi 2D electronic structure is seen in the band structure where the dispersion along the $c^*$ direction ($\Gamma$--Z, B--$\Delta$, and Y--T) is relatively flatter than the dispersion in the $a^*$--$b^*$ plane near $E_\mathrm{F}$. 

\begin{figure}[tb]
\begin{center}
\includegraphics[width=0.58\textwidth]{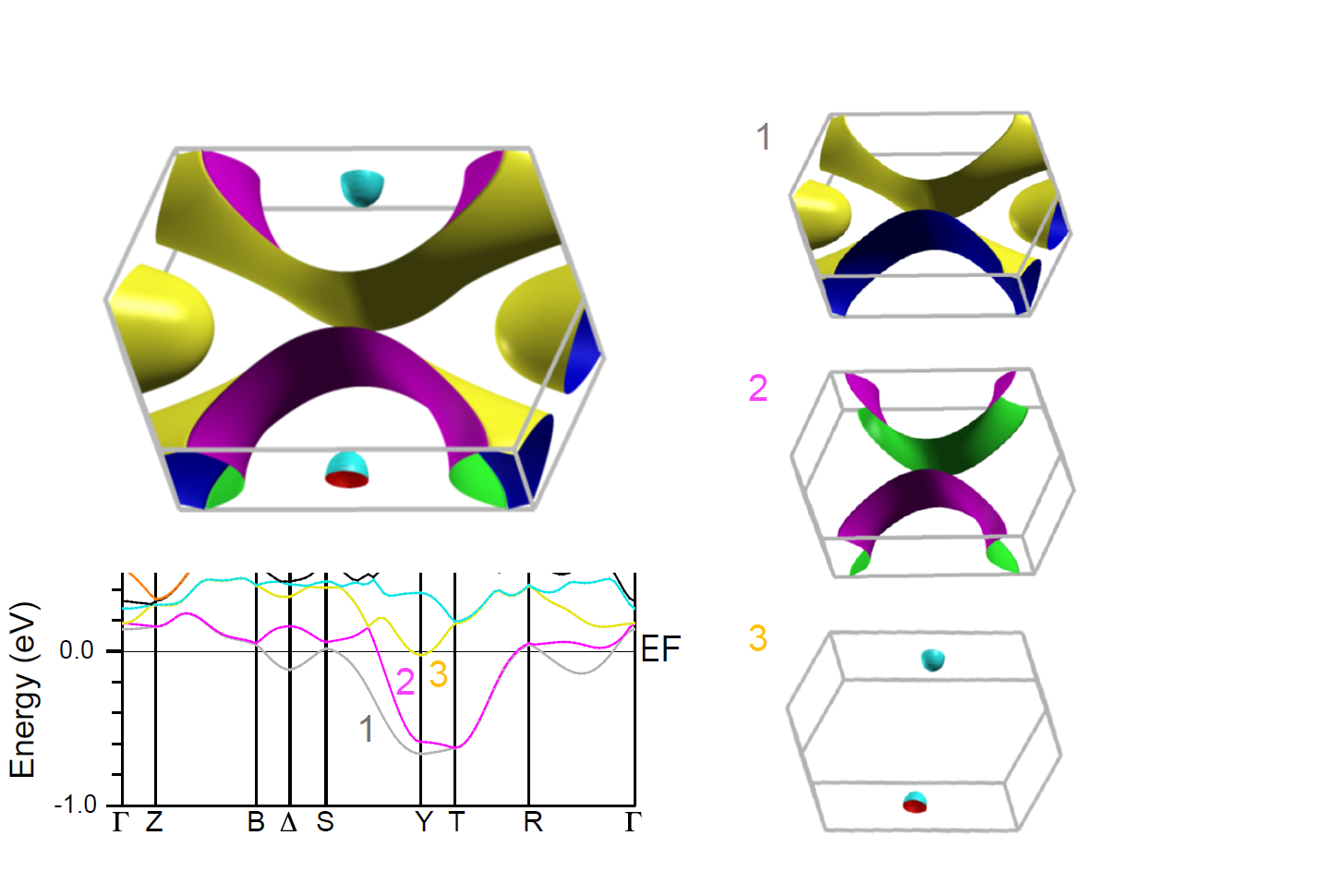}
\caption{Fermi surfaces (left top) of AlTi$_2$O$_5$ and band structure near $E_\mathrm{F}$ (left bottom). There are three band that cross at $E_\mathrm{F}$ that are numbered as 1, 2, and 3. The Fermi surface of each band is shown in the right three panels.}
\label{figFS}
\end{center}
\end{figure}

There are three bands that cross at $E_\mathrm{F}$ as numbered in Fig.~\ref{figFS}. The Fermi surfaces are shown in Fig.~\ref{figFS}. The band 2 shows a cylindrical shape whose central axis is along the Y--T line, reflecting the quasi 2D electronic structure. The band 1 also gives a similar cylindrical shape together with an electron pocket centered at the $\Delta$ point.   The band 3 shows a small pocket centered at the Y point. The presence of the cylindrical Fermi surfaces indicates a possible nesting-driven order as is the case of, for example, iron-pnictide superconductors.  In the next section, we will study such a possibility and propose a nesting-driven CDW order.    
 
\subsection{Randomness: Super-cell calculation}
\label{Sec2-2}

As shown in Table~\ref{table1}, the experimental analysis of the crystal structure in AlTi$_2$O$_5$ suggests random distribution of Al and Ti in the A and B sites. Based on the analysis, we consider two cases of occupation ratio of $\mathrm{Al}:\mathrm{Ti}$ for the A and B sites: the case (a) with $2:8$ for the A site and $4:6$ for the B site, and the case (b) with $3:7$ for the A site and $3.5:6.5$ for the B site. 

In order to clarify the effect of the randomness on the electronic structure, we performed super-cell calculations of DOS with $2\times 2\times 2$ unit cells employing the Vienna {\it Ab-initio} Simulation Package (VASP)~\cite{Kresse1996} using the projector augmented wave pseudopotentials~\cite{Kresse1999}. We use GGA using the PBE exchange correlation potential~\cite{Perdew1996}. An $8\times 8 \times 8$ $k$-point grid is adopted for the self-consistent calculations for DOS.

\begin{figure}[tb]
\begin{center}
\includegraphics[width=0.48\textwidth]{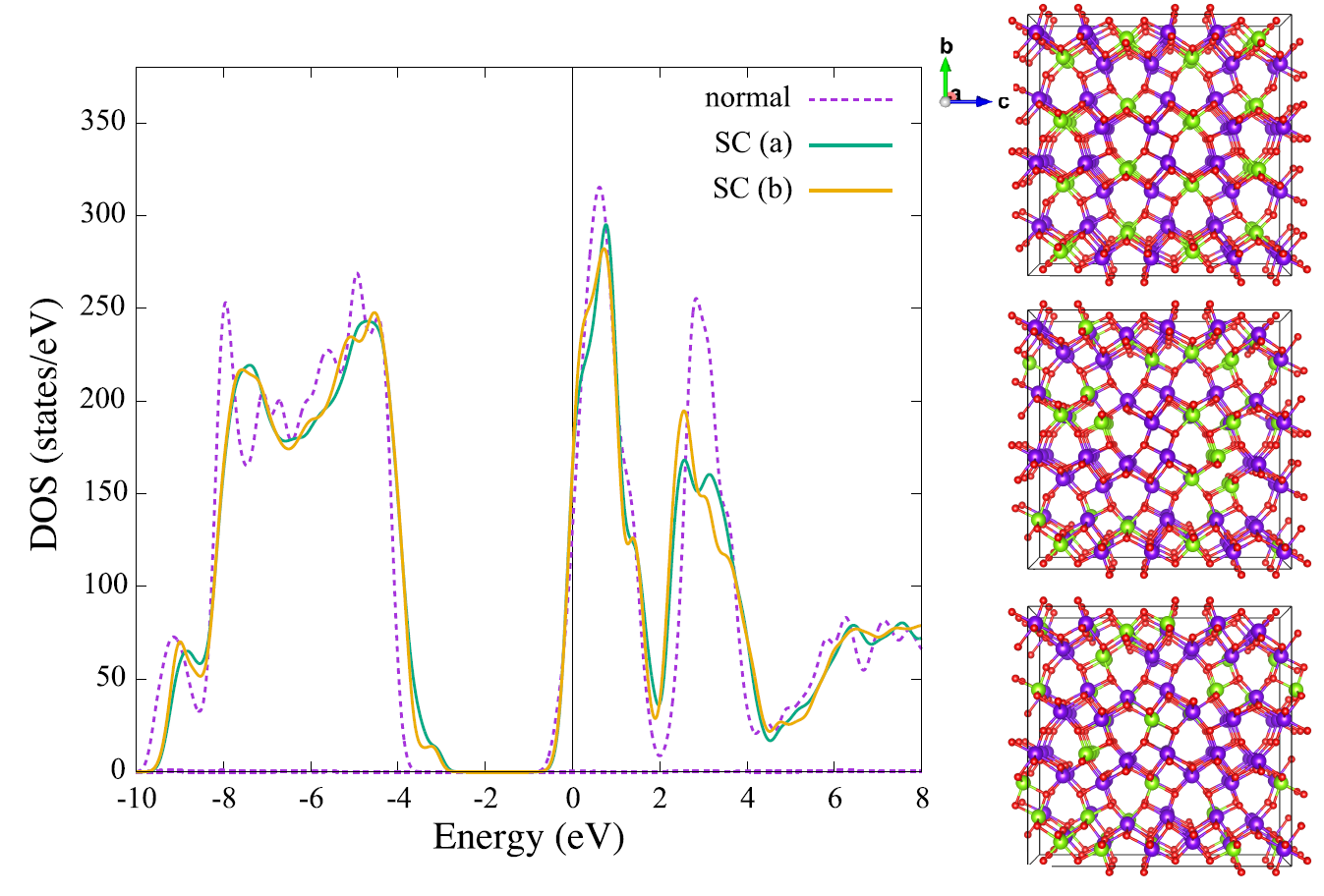}
\caption{Calculated total DOS for AlTi$_2$O$_5$ using the $2\times 2\times 2$ super cell. The purple dotted line represents the normal case without randomness. The green thick solid and orange thin solid lines represent the case (a) [SC(a)] and the case (b) [SC(b)] of the super cell, respectively, with occupation ratio of Al and Ti (see text). The right-top, right-middle, and right-bottom panels denote the super cell for the normal, SC(a), and SC(b), respectively. The purple, green, and red circles correspond to Ti, Al, and O atoms, respectively.}
\label{figSC}
\end{center}
\end{figure}

Figure~\ref{figSC} shows the total DOS in the $2\times 2\times 2$ cells. The dotted line represents the normal case without randomness whose structure is shown in the right-top panel in Fig.~\ref{figSC}. Since the number of $k$-point grid in the super-cell calculations is smaller than the primitive-cell calculation shown in Fig.~\ref{figBS}, fine structures such as a peak at 0.15~eV in Fig.~\ref{figBS} are smeared. However, a broad peak at 0.6~eV and a dip at 2~eV are consistent with the DOS in Fig.~\ref{figBS}. Therefore, the present super-cell calculation captures global behaviors in DOS of nonrandom AlTi$_2$O$_5$.  The total DOS for the case (a) and the case (b) of the occupation ratio is shown by the green thick solid line and the orange thin solid line, respectively, in Fig.~\ref{figSC}. The random distribution of Al and Ti in the $2\times 2\times 2$ unit cells for the case (a) and the case (b) are shown in the right-middle and right-bottom panels, respectively. Introducing such randomness causes the broadening of large peaks. However, the DOS at $E_\mathrm{F}$ hardly changes, remaining as a metal even in the presence of the randomness of Al and Ti. Therefore, it is clear that the randomness cannot be the origin of insulating behavior in AlTi$_2$O$_5$.

\section{Spin and charge susceptibility}
\label{Sec4}
In order to take a crucial insight for the origin of insulating behavior in AlTi$_2$O$_5$, we focus on the nonrandom AlTi$_2$O$_5$ and calculate spin and charge susceptibilities using RPA for the tight-binding model constructing from the first-principles calculation.
\subsection{Formulation}
\label{Sec4-1}
In order to examine spin and charge susceptibilities in RPA, we construct the tight-binding Hamiltonian $H_0$ composed of Ti $3d$ orbitals, which is written as
\begin{align}
H_0 = \sum_{i,j} \sum_{\rho, \rho'} \sum_{\mu, \nu} \sum_\sigma  t_{i\rho j\rho'}^{\mu\nu}  c_{i\rho,\mu,\sigma}^\dagger c_{j\rho',\nu,\sigma},
\label{H0}
\end{align}
where $c_{i\rho, \mu, \sigma}^\dagger$ creates an electron with orbital $\mu$ and spin $\sigma$ at the position $\mathbf{R}_i+\mathbf{r}_\rho$ for sublattice $\rho$ in unit cell $i$. The matrix element $t_{i\rho j\rho'}^{\mu\nu}$ represents the hopping of an electron between the $\mu$ orbital at sublattice $\rho$ in unit cell $i$ and the $\nu$ orbital at sublattice $\rho'$ in unit cell $j$. The values of $t_{i\rho j\rho}^{\mu\nu}$ are obtained by using an interface from LAPW to maximally localized Wannier function~\cite{wien2wannier} and Wannier90~\cite{wannier90}, inputting the data from the GGA band structure without randomness mentioned above. We set 20 orbitals (five $3d$ orbitals in four Ti atoms in the unit cell) in $H_0$ and use $t_{ij}^{\mu\nu}$ that perfectly reproduces 12 bands in the region from $-$0.8~eV to 1.8~eV in Fig.~\ref{figBS}.

The Coulomb interactions for the $3d$ orbitals are given by~\cite{Oles1983,Sugimoto2013}
\begin{align}
H_{\mathrm{int}} =& \frac{U}{2} \sum_{i,\rho} \sum_{\mu, \sigma} c_{i\rho, \mu, \sigma}^\dagger c_{i\rho, \mu, \sigma} c_{i\rho, \mu, -\sigma}^\dagger c_{i\rho, \mu, -\sigma} \notag \\
 &+ \frac{U-2J}{2} \sum_{i,\rho} \sum_{\mu \neq \nu, \sigma} c_{i\rho, \mu, \sigma}^\dagger c_{i\rho, \mu, \sigma} c_{i\rho, \nu, -\sigma}^\dagger c_{i\rho, \nu, -\sigma}  \notag \\
 &+ \frac{U-3J}{2} \sum_{i, \rho} \sum_{\mu \neq \nu, \sigma}  c_{i\rho, \mu, \sigma}^\dagger c_{i\rho, \mu, \sigma} c_{i\rho, \nu, \sigma}^\dagger c_{i\rho, \nu, \sigma} \notag \\
  &- \frac{J}{2} \sum_{i, \rho} \sum_{\mu \neq \nu, \sigma} c_{i\rho, \mu, \sigma}^\dagger c_{i\rho, \mu, -\sigma} c_{i\rho, \nu,  -\sigma}^\dagger c_{i\rho, \nu, \sigma}  \notag \\
 &+  \frac{J}{2} \sum_{i, \rho} \sum_{\mu \neq \nu, \sigma} c_{i\rho, \mu, \sigma}^\dagger c_{i\rho, \nu, \sigma} c_{i\rho, \mu, -\sigma}^\dagger c_{i\rho, \nu, -\sigma},
\label{Hint}
\end{align}
where $U$ and $J$ are on-site Coulomb and exchange interaction for $3d$ electrons, respectively.

Introducing the Fourier transformation of $c_{i\rho, \mu, \sigma}$ to the momentum space, $c_{\mathbf{k}, \rho, \mu, \sigma}=N^{-1/2} \sum_{i} \exp^{ -\i \mathbf{k} \cdot (\mathbf{R}_i+\mathbf{r}_\rho)} c_{i\rho, \mu, \sigma}$, we define the orbital component of the static susceptibility $\chi^{\sigma\sigma'}(\mathbf{q})$ as
\begin{align}
\left(\chi^{\sigma\sigma'}\right)_{\kappa\lambda;\mu\nu}^{\rho;\rho'} (\mathbf{q})
 =& \frac{\i}{N} \sum_{\mathbf{k}, \mathbf{k}'} \int^\infty_0 dt \langle [c_{\mathbf{k}, \rho,\lambda, \sigma}^\dagger(t)   c_{\mathbf{k} + \mathbf{q}, \rho, \kappa, \sigma}(t), \notag \\
& c_{\mathbf{k}' + \mathbf{q}, \rho', \nu, \sigma'}^\dagger  c_{\mathbf{k}', \rho', \mu, \sigma'} ] \rangle,
\label{chi}
\end{align}
where $N$ is the total number of sites, $\langle[\cdots ,\cdots]\rangle$ is the average value of the anticommutator, $c_{\mathbf{k}, \rho, \mu, \sigma}(t)$ is the Heisenberg representation of $c_{\mathbf{k}, \rho, \mu, \sigma}$. We define the spin susceptibility $\chi^s(\mathbf{q})$ and charge susceptibility $\chi^c(\mathbf{q})$ in the paramagnetic phase as
\begin{align}
\chi^{s}(\mathbf{q})=\frac{1}{2}\sum_\sigma\left(\chi^{\sigma\sigma}(\mathbf{q})-\chi^{\sigma\bar{\sigma}}(\mathbf{q})\right)
\label{chis} 
\end{align}
and
\begin{align}
\chi^{c}(\mathbf{q})=\frac{1}{4}\sum_\sigma\left(\chi^{\sigma\sigma}(\mathbf{q})+\chi^{\sigma\bar{\sigma}}(\mathbf{q})\right)
\label{chic} 
\end{align}
respectively, where $\bar{\sigma}$ denotes $\downarrow$ ($\uparrow$) for $\sigma=\uparrow$ ($\downarrow$).

Within the multi-orbital RPA~\cite{Sugimoto2013}, we have
\begin{align}
 \begin{pmatrix}
	\chi^{\uparrow\uparrow} \\ \chi^{\downarrow \uparrow}
 \end{pmatrix}
=
 \begin{pmatrix}
	\chi_0 \\ 0
 \end{pmatrix} 
+
 \begin{pmatrix}
	\chi_0 V^{\uparrow \uparrow} & \chi_0 V^{\uparrow \downarrow} \\
	\chi_0 V^{\downarrow\uparrow} & \chi_0 V^{\downarrow\downarrow}
 \end{pmatrix}
 \begin{pmatrix}
	\chi^{\uparrow\uparrow} \\ \chi^{\downarrow \uparrow}
 \end{pmatrix},
\label{chiV}
\end{align}
where the bare susceptibility $\chi_0$ is given by
\begin{align}
 &\left(\chi_0\right)_{\kappa\lambda;\mu\nu}^{\rho;\rho'} (\mathbf{q}) \notag \\
 &= -\frac{1}{N} \sum_{\mathbf{k}, \alpha, \alpha'} \frac{ f(\varepsilon_{\mathbf{k}+\bm{q};\alpha}) - f(\varepsilon_{\mathbf{k};\alpha'}) }{ \varepsilon_{\mathbf{k}+\mathbf{q};\alpha} - \varepsilon_{\mathbf{k};\alpha'} - i \eta} \notag \\
& \times
\psi_{\rho, \kappa; \alpha} (\mathbf{k} + \mathbf{q})
\psi_{\rho', \nu; \alpha}^* (\mathbf{k} + \mathbf{q})
\psi_{\rho, \lambda; \alpha'}^* (\mathbf{k})
\psi_{\rho', \mu; \alpha'} (\mathbf{k}),
\label{chi0}
\end{align}
with the wavefunction $\psi_{\rho, \mu; \alpha} (\mathbf{k})$ and eigenvalue $\varepsilon_{\mathbf{k};\alpha}$ for the momentum $\mathbf{k}$ and band $\alpha$.  The product of $\chi_0$, $\chi^{\sigma\sigma'}$, and the interaction matrix $V^{\sigma_1\sigma_2}$ in (\ref{chiV}) is taken as
\begin{align}
\left(\chi_0 V^{\sigma_1\sigma_2} \chi^{\sigma_3\sigma_4} \right)_{\kappa\lambda;\mu\nu}^{\rho;\rho'}(\mathbf{q}) 
=& \sum_{\kappa', \lambda', \mu', \nu'} \left(\chi_0\right)_{\kappa\lambda;\mu'\nu'}^{\rho;\rho'}(\mathbf{q}) \notag \\ 
\times \left(V^{\sigma_1\sigma_2}\right)_{\nu'\mu';\lambda'\kappa'} & \left(\chi^{\sigma_3\sigma_4} \right)_{\kappa'\lambda';\mu\nu}^{\rho;\rho'}(\mathbf{q}),
\label{cVc}
\end{align}
where the nonzero elements of $V^{\sigma\sigma}$ are given by $-U+3J$ ($U-3J$) for $\mu=\nu\neq\kappa=\lambda$ ($\mu=\lambda\neq\nu=\kappa$) and those of $V^{\sigma\bar{\sigma}}$ are given by $-U$ ($-U+2J$) for $\mu=\nu=\kappa=\lambda$ ($\mu=\nu\neq\kappa=\lambda$) and $-J$ for $\mu=\kappa\neq\nu=\lambda$ and $\mu=\lambda\neq\nu=\kappa$. 

To extract dominant fluctuations in the spin and charge degrees of freedom, we use an eigenmode analysis~\cite{Uehara2015} of $\chi^{s}(\mathbf{q})$ and $\chi^{c}(\mathbf{q})$. In this analysis, we diagonalize the matrix form of $\chi^{s}(\mathbf{q})$ and $\chi^{c}(\mathbf{q})$ for each $\mathbf{q}$, whose matrix size is $100\times 100$ because of four sublattices with five orbitals giving $4\times5^2=100$, and examine the largest eigenvalue and its eigenvector: the eigenvalue corresponds to the amplitude of the fluctuation and the eigenvector is regarded as the direction of the fluctuation~\cite{Uehara2015}.

In the calculation of $\chi_0(\mathbf{q})$, we sum over the first BZ using 12$^3$ grid points. The momentum transfer $\mathbf{q}$ is defined as $\mathbf{q}=q_1\mathbf{b}_1+q_2\mathbf{b}_2+q_3\mathbf{b}_3$ with $\mathbf{b}_1=(2\pi/a,2\pi/b,0)$, $\mathbf{b}_2=(-2\pi/a,2\pi/b,0)$, and $\mathbf{b}_3=(0,0,2\pi/c)$. We consider zero temperature and set $\eta = 10^{-5}$~eV in Eq.~(\ref{chi0}). For the value of $U$ and $J$ in Eq.~(\ref{Hint}), we chose two cases of the ratio $J/U$, $J/U=0.15$ and $J/U=0.3$, since we do not know its precise value. For each $J/U$, we increase $U$ from zero and examine the case where strong enhancement of the largest eigenevalue in either $\chi^{s}(\mathbf{q})$ or $\chi^{c}(\mathbf{q})$ appears for a certain $\mathbf{q}$.  We find that a diverging behavior emerges in $\chi^{c}(\mathbf{q})$ at $U=0.2$~eV (2.0~eV) for $J/U=0.15$ (0.3) inside the $q_3=0$ ($q_3=\pi$) plane, as shown below.  

\subsection{Calculated results}
\label{Sec4-2}

\begin{figure}[tb]
\begin{center}
\includegraphics[width=0.65\textwidth]{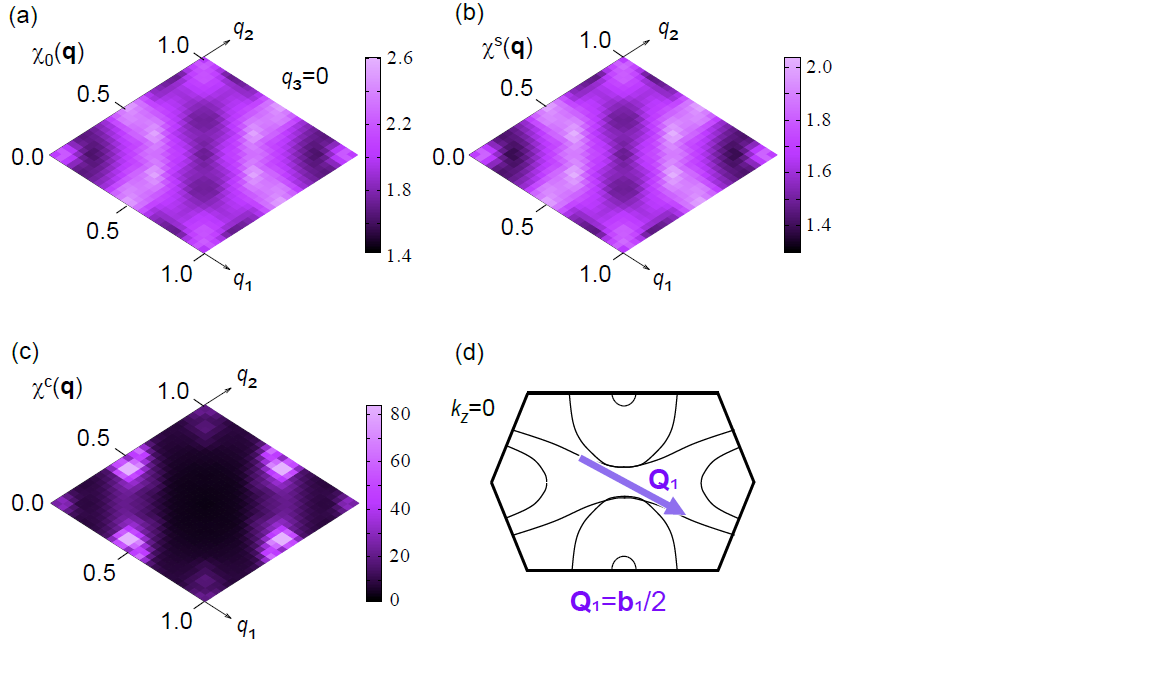}
\caption{Contour plot of the largest eigenvalues of (a) $\chi_0(\mathbf{q})$, (b) $\chi^s(\mathbf{q})$, and (c) $\chi^c(\mathbf{q})$ in the $q_3=0$ plane for $J/U=0.15$ and $U=0.2$~eV.  (d) Fermi surfaces in the $k_z=0$ plane. The vector $\mathbf{Q}_1$ is a possible nesting vector.}
\label{figSU1}
\end{center}
\end{figure}

Figure~\ref{figSU1}(a) shows the contour plot of the largest eigenevalue of $\chi_0(\mathbf{q})$ in the $q_3=0$ plane for $J/U=0.15$ and $U=0.2$~eV. 
There is a broad region showing large values along the arc connecting $(q_1,q_2)=(0.5,0.0)$ and $(0.0, 0.5)$ and its symmetric region.  These large values come from nesting properties with vectors connecting two Fermi surfaces approximately. Figure~\ref{figSU1}(d) shows the Fermi surfaces on the $k_z=0$ plane and a possible nesting vector $\mathbf{Q}_1=\mathbf{b}_1/2$ though the nesting is imperfect. The largest eigenvalues of $\chi^s(\mathbf{q})$ do not change much from those of $\chi_0(\mathbf{q})$ as shown in Fig.~\ref{figSU1}(b). This is consistent with the fact that there is no magnetic solution in the first-principles calculations as mentioned above. On the other hand, the largest eigenvalues of $\chi^c(\mathbf{q})$ exhibit a diverging behavior near $\mathbf{Q}_1$ and its symmetric points as shown in Fig.~\ref{figSU1}(c). These points are close to the arc regions seen in $\chi_0(\mathbf{q})$. Therefore, the origin of the diverging behavior might be related to the tendency toward nesting of the Fermi surfaces. 

\begin{figure}[tb]
\begin{center}
\includegraphics[width=0.65\textwidth]{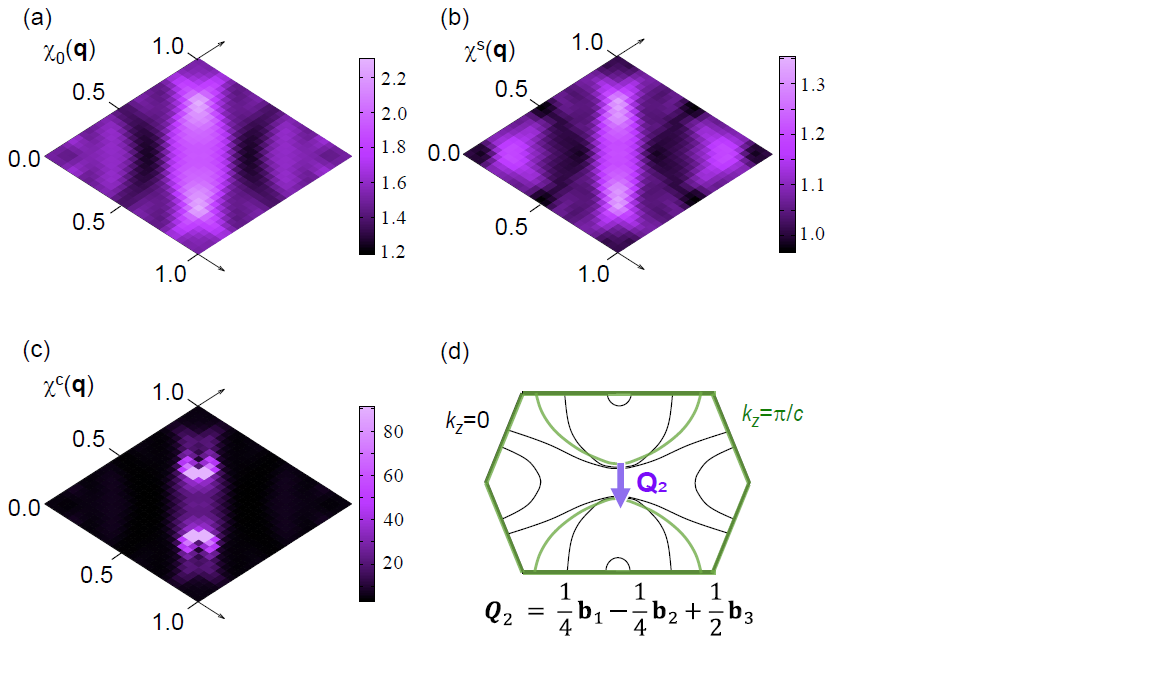}
\caption{Contour plot of the largest eigenvalues of (a) $\chi_0(\mathbf{q})$, (b) $\chi^s(\mathbf{q})$, and (c) $\chi^c(\mathbf{q})$ in the $q_3=1$ plane for $J/U=0.3$ and $U=2.0$~eV.  (d) Fermi surfaces in the $k_z=0$ plane (black lines) and in the $k_z=\pi/c$ (green lines). The vector $\mathbf{Q}_2$ is a possible nesting vector. }
\label{figSU2}
\end{center}
\end{figure}

Figure~\ref{figSU2}(a) shows the contour plot of the largest eigenevalue of $\chi_0(\mathbf{q})$ in the $q_3=1$ plane for $J/U=0.3$ and $U=2.0$~eV. In this case, nesting properties appear along the line from $(q_1,q_2)=(1,0)$ to $(0,1)$ and the largest signal emerges close to $\mathbf{Q}_2=\mathbf{b}_1/4-\mathbf{b}_2/4+\mathbf{b}_3/2$ corresponding to $(q_1,q_2,q_3)=(1/4,-1/4,1/2)=(3/4,1/4,1/2)$, which is a possible nesting vector connecting the Fermi surfaces between the $k_z=0$ and $k_z=\pi/c$ planes as shown in Fig.~\ref{figSU2}(d).  The largest eigenvalues of $\chi^s(\mathbf{q})$ do not change much from those of $\chi_0(\mathbf{q})$ as shown in Fig.~\ref{figSU2}(b). This is again consistent with the fact that there is no magnetic solution in the first-principles calculations. On the other hand, the largest eigenvalues of $\chi^c(\mathbf{q})$ exhibit a diverging behavior near $\mathbf{Q}_2$ and its symmetric points as shown in Fig.~\ref{figSU2}(c). Therefore, the origin of the diverging behavior is also related to the tendency toward nesting of the Fermi surfaces. 

In the single-band Hubbard model, it is impossible to obtain the CDW instability within RPA as long as the effective interaction is repulsive. However, in the multi-orbital Hubbard model, the interaction matrix elements in Eq.~(\ref{cVc}) may cause complicated orbital-dependent behaviors and consequently lead to a tendency toward CDW instability. The tendency will be enhanced by not only the effect of the nesting but also van Hove singularities near $E_\mathrm{F}$ as discussed for the CDW in NdSe$_2$~\cite{Sadowski2013}. The $E_\mathrm{F}$ in AlTi$_2$O$_5$ is actually located just below a van Hove singularity as shown in Fig.~\ref{figBS}. Therefore, the CDW instability in AlTi$_2$O$_5$ may be related to not only the enhancement of $\chi_0(\mathbf{q})$ but also the closeness of $E_\mathrm{F}$ to the van Hove singularity.   

\begin{figure}[tb]
\begin{center}
\includegraphics[width=0.8\textwidth]{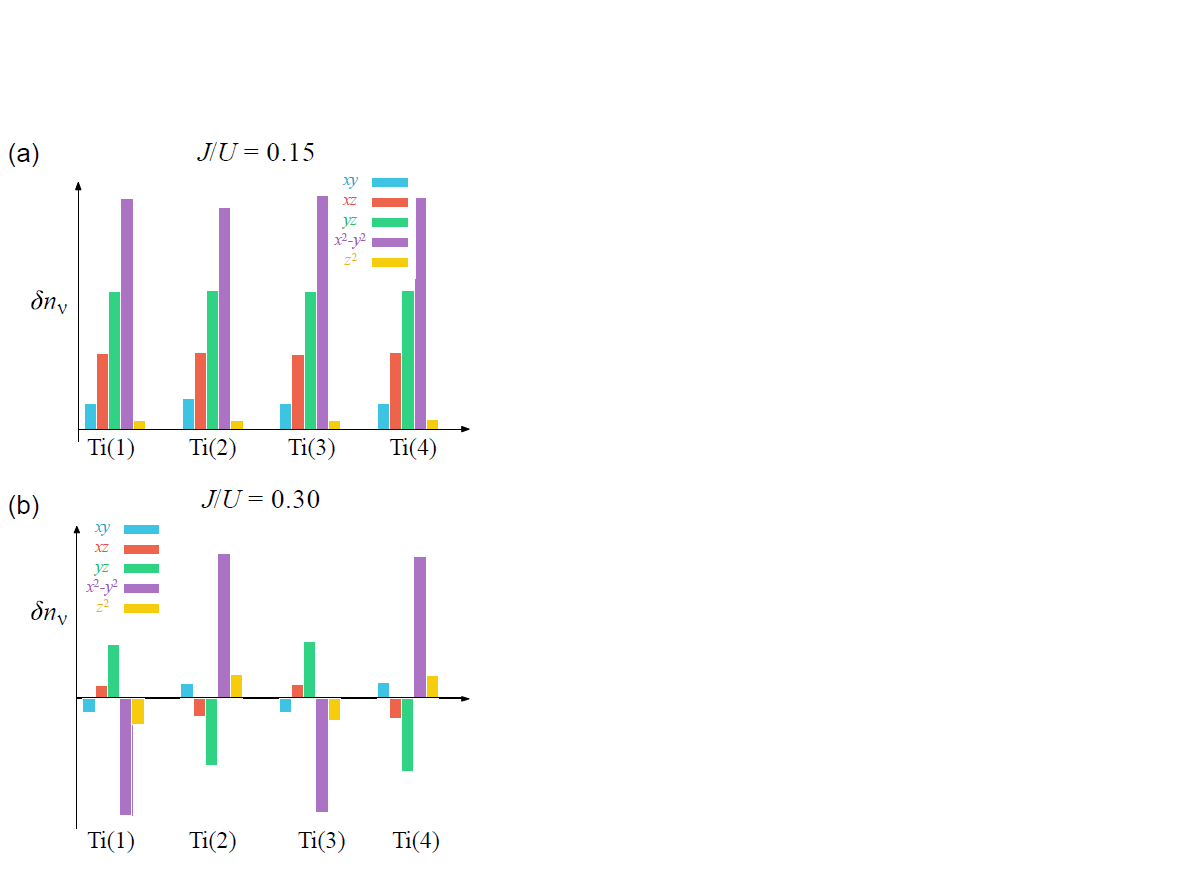}
\caption{Histogram of orbital-dependent charge fluctuation $\delta n_\nu$ with $\nu=x^2-y^2$, $yz$, and $xz$ for $t_{2g}$ orbitals and $\nu=yz$ and $3z^2-y^2$ ($z^2$) for $e_g$ orbitals. (a) $J/U=0.15$ and (b) $J/U=0.3$. $\delta n_\nu$ is obtained from the eigenvector of $\chi^c(\mathbf{q})$ at the $\mathbf{q}$ point exhibiting the largest eigenvalue in Figs.~\ref{figSU1}(c) and \ref{figSU2}(c).}
\label{figDN}
\end{center}
\end{figure}

In order to see orbital-dependent charge fluctuation $\delta n_\nu$ in each Ti site, we examine the eigenvector of $\chi^c(\mathbf{q})$ at the $\mathbf{q}$ point exhibiting the largest eigenvalue.  $\delta n_\nu$ is plotted as the form of histogram with arbitrary unit in Figs.~\ref{figDN}(a) and \ref{figDN}(b) for $J/U=0.15$ and 0.3, respectively. For $J/U=0.15$, dominant charge fluctuations for each Ti site are organized by $\nu=x^2-y^2$, $yz$ and $xz$, all of which consist of $t_{2g}$ orbitals.  Since $\delta n_\nu$ in each site has the same fluctuations, the eigenmode of the largest $\chi^c(\mathbf{q})$ can be regarded as an acoustic mode (in-phase mode). On the other hand, $\delta n_\nu$ for $J/U=0.3$ in Fig.~\ref{figDN}(b) shows opposite directions between Ti(1) [Ti(3)] and Ti(2) [Ti(4)], indicating an optical mode (out-of-phase mode). Again, $\nu=x^2-y^2$ has the largest fluctuation in every sites and $\nu=yz$ has the secondly largest fluctuation. 

Since $\mathbf{Q}_1$ and $\mathbf{Q}_2$ in Fig.~\ref{figSU1} and Fig.~\ref{figSU2}, respectively, are commensurate with lattice spacing, it is expected that electron-phonon coupling will induce lattice distortion, leading to a CDW state with insulating properties. Taking into account the $\mathbf{Q}_1$ vector, we may predict a stripe-type CDW state with alternating Ti$^{3+}$ with a $3d$ electron in predominately $x^2-y^2$ orbital and Ti$^{4+}$ with no $3d$ electron for $J/U=0.15$. This is similar to the low-temperature phase of Ti$_4$O$_7$~\cite{Marezio1973}. On the other hand,  for $J/U=0.3$, $\mathbf{Q}_2$ indicates a longer-range period of CDW. If the insulating behavior in AlTi$_2$O$_5$ comes from CDW, it is important to identify characteristic wave vector and orbitals contributing to the order. This remains as a challenging experimental problem to be solved in the future.

\section{Summary}
\label{Sec5}
We synthesized a new titanium-oxide compound AlTi$_2$O$_5$, whose formal valence of Ti is $3.5+$ that is the same as Ti$_4$O$_7$. In the synthesized sample, we found that Al and Ti are randomly distributed on the A and B sites, although more than 60~\% of the B sites are occupied by Ti. AlTi$_2$O$_5$ is insulating, showing a huge resistivity enhancement below 120 K while it slightly increases with decreasing temperature down to about 120~K. Using the determined atomic coordinates, we performed the first-principles band-structure calculations. We found cylindrical Fermi surfaces due to an approximate layered structure in the case of nonrandom distribution of Al and Ti. The Fermi surfaces indicate a possible nesting-driven order. We found neither the insulating nor magnetic ground state even for a moderate value of $U$. We also examined the effect of randomness in Al and Ti on the DOS, and it found to be not the origin of insulating behavior in AlTi$_2$O$_5$.  

In order to take a crucial insight for the origin of insulating behavior in AlTi2O5, we focused on the nonrandom
AlTi$_2$O$_5$. After constructing a tight-binding model from the first-principles calculation, we performed the calculation of spin and charge susceptibilities using RPA. We found that the charge susceptibility shows strong enhancement at the wave vectors close to nesting conditions. Their positions depend on the choice of Coulomb interactions. One of the wave vectors corresponds to a CDW state whose charge ordering is similar to the low-temperature phase of Ti$_4$O$_7$. The enhancement is expected to be due to nesting properties of the Fermi surfaces in addition to the presence of the van Hove singularity near the Fermi level. These results give a hint for the mechanism of insulating behavior and contribute to further experimental effort to characterize the physical properties of AlTi$_2$O$_5$.     
\\

\begin{acknowledgments}
We thank Y. Sakai for his contribution in the initial stage of theoretical part. This work was supported by MEXT, Japan, as a social and scientific priority issue (creation of new functional devices and high-performance materials to support next-generation industries) to be tackled by using a post-K computer. The numerical calculation was partly carried out at the facilities of the Supercomputer Center, Institute for Solid State Physics, University of Tokyo.).
\end{acknowledgments}



\end{document}